\documentclass[11pt]{article}
\usepackage{graphicx}

\catcode`\@=11
\long\def\@makefntext#1{
\protect\noindent \hbox to 3.2pt {\hskip-.9pt
$^{{\eightrm\@thefnmark}}$\hfil}#1\hfill}       %CAN BE USED

\def\@makefnmark{\hbox to 0pt{$^{\@thefnmark}$\hss}}    %ORIGINAL

\def\ps@myheadings{\let\@mkboth\@gobbletwo
\def\@oddhead{\hbox{}
\rightmark\hfil\eightrm\thepage}
\def\@oddfoot{}\def\@evenhead{\eightrm\thepage\hfil
\leftmark\hbox{}}\def\@evenfoot{}
\def\sectionmark##1{}\def\subsectionmark##1{}}

\oddsidemargin=\evensidemargin
\newcounter{sectionc}\newcounter{subsectionc}\newcounter{subsubsectionc}
\renewcommand{\section}[1] {\vspace{12pt}\addtocounter{sectionc}{1}
\setcounter{subsectionc}{0}\setcounter{subsubsectionc}{0}\noindent
    {\tenbf\thesectionc. #1}\par\vspace{5pt}}
\renewcommand{\subsection}[1] {\vspace{12pt}\addtocounter{subsectionc}{1}
    \setcounter{subsubsectionc}{0}\noindent
    {\bf\thesectionc.\thesubsectionc. {\kern1pt \bfit #1}}\par\vspace{5pt}}
\renewcommand{\subsubsection}[1] {\vspace{12pt}\addtocounter{subsubsectionc}{1}
    \noindent{\tenrm\thesectionc.\thesubsectionc.\thesubsubsectionc.
    {\kern1pt \tenit #1}}\par\vspace{5pt}}

\newcommand{\textlineskip}{\baselineskip=13pt}

\def\eightcirc{
\begin{picture}(0,0)
\put(4.4,1.8){\circle{6.5}}
\end{picture}}
\def\eightcopyright{\eightcirc\kern2.7pt\hbox{\eightrm c}}

\def\abstracts#1#2#3{{
    \centering{\begin{minipage}{4.5in}\baselineskip=10pt\footnotesize
    \parindent=0pt #1\par
    \parindent=15pt #2\par
    \parindent=15pt #3
    \end{minipage}}\par}}

\renewenvironment{thebibliography}[1]
    {\frenchspacing
     \ninerm\baselineskip=11pt
     \begin{list}{\arabic{enumi}.}
        {\usecounter{enumi}\setlength{\parsep}{0pt}
     \setlength{\leftmargin 12.7pt}{\rightmargin 0pt} %FOR 1--9 ITEMS
         \setlength{\itemsep}{0pt} \settowidth
    {\labelwidth}{#1.}\sloppy}}{\end{list}}

\newcounter{itemlistc}
\newcounter{romanlistc}
\newcounter{alphlistc}
\newcounter{arabiclistc}

\def\@citex[#1]#2{\if@filesw\immediate\write\@auxout
    {\string\citation{#2}}\fi
\def\@citea{}\@cite{\@for\@citeb:=#2\do
    {\@citea\def\@citea{,}\@ifundefined
    {b@\@citeb}{{\bf ?}\@warning
    {Citation `\@citeb' on page \thepage \space undefined}}
    {\csname b@\@citeb\endcsname}}}{#1}}

\newif\if@cghi
\def\cite{\@cghitrue\@ifnextchar [{\@tempswatrue
    \@citex}{\@tempswafalse\@citex[]}}
\def\citelow{\@cghifalse\@ifnextchar [{\@tempswatrue
    \@citex}{\@tempswafalse\@citex[]}}
\def\@cite#1#2{{$\null^{#1}$\if@tempswa\typeout
    {IJCGA warning: optional citation argument
    ignored: `#2'} \fi}}

\def\@refcitex[#1]#2{\if@filesw\immediate\write\@auxout
    {\string\citation{#2}}\fi
\def\@citea{}\@refcite{\@for\@citeb:=#2\do
    {\@citea\def\@citea{, }\@ifundefined
    {b@\@citeb}{{\bf ?}\@warning
    {Citation `\@citeb' on page \thepage \space undefined}}
    \hbox{\csname b@\@citeb\endcsname}}}{#1}}

\def\@refcite#1#2{{#1\if@tempswa\typeout
        {IJCGA warning: optional citation argument
    ignored: `#2'} \fi}}

\def\refcite{\@ifnextchar[{\@tempswatrue
    \@refcitex}{\@tempswafalse\@refcitex[]}}

\def\pmb#1{\setbox0=\hbox{#1}
    \kern-.025em\copy0\kern-\wd0
    \kern.05em\copy0\kern-\wd0
    \kern-.025em\raise.0433em\box0}

\def\fnt#1#2{\footnotetext{\kern-.3em
    {$^{\mbox{\scriptsize #1}}$}{#2}}}

\def\runninghead#1#2{\pagestyle{myheadings}
\markboth{{\protect\footnotesize\it{\quad #1}}\hfill}
{\hfill{\protect\footnotesize\it{#2\quad}}}}
\font\tenrm=cmr10
\font\tenit=cmti10
\font\tenbf=cmbx10
\font\bfit=cmbxti10 at 10pt
\font\ninerm=cmr9

\font\eightrm=cmr8

\def\qed{\hbox{${\vcenter{\vbox{            %HOLLOW SQUARE
   \hrule height 0.4pt\hbox{\vrule width 0.4pt height 6pt
   \kern5pt\vrule width 0.4pt}\hrule height 0.4pt}}}$}}

\begin{document}

\newpage

\runninghead{} {Supersymmetry of FRW barotropic cosmologies}

\normalsize\textlineskip
\thispagestyle{empty}
\setcounter{page}{1}

%\copyrightheading{}                     %{Vol. 0, No.0 (1992) 000--000}

\vspace*{0.88truein}

%\fpage{1} %%%%%%%%%%%%%%%%%%%%%%%%%%%%%%%%%%%%%%%%%%%%%%%%%%%%%%%%%%%
%\centerline{Mod. Phys. Lett. A xx (2000) i-f
%[quant-ph/0003xxx v3]}
%\centerline{\footnotesize\it To understand hydrogen is to understand all of
%physics}
%\centerline{\footnotesize in ``The Yin and Yang of Hydrogen", D. Kleppner,
%Phys. Today, April 1999, pp. 11-12}
\bigskip
%\bigskip
\centerline{\bf SUPERSYMMETRY OF FRW BAROTROPIC COSMOLOGIES}
\vspace*{0.035truein}
\vspace*{0.37truein}
\vspace*{10pt}
\centerline{\footnotesize H.C. ROSU\footnote{E-mail: \tt hcr@ipicyt.edu.mx \hfill hcrcos1.tex} $\,$ %\hfill  hcrcos.tex} $\,$
 and P. OJEDA-MAY\footnote{E-mail: \tt pedro@ipicyt.edu.mx}}
\vspace*{0.015truein}
%\centerline{\footnotesize [Received 17 August 1999] }
%\baselineskip=10pt
\centerline{\footnotesize Potosinian Institute of Science and
Technology ({\em IPICyT})} \centerline{ \footnotesize Apartado
Postal 3-74 Tangamanga, 78231 San Luis Potos\'{\i}, Mexico}
\vspace*{0.225truein}
%%%%%%%%%%%%%%%%%%%%%%%%%%%%%%%%%%%%%%%%%%%%%
%\publisher{(June 2002) - Mod Phys. Lett. A}
%{(May 14, 2000)}
%\centerline{\footnotesize Dated Feb 25th 2004 $\qquad$ file: DIRACC.TEX}
%%%%%%%%%%%%%%%%%%%%%%%%%%%
\vspace*{0.21truein} \abstracts{Barotropic FRW cosmologies are
presented from the standpoint of nonrelativistic supersymmetry.
First, we reduce the barotropic FRW system of differential equations
to simple harmonic oscillator differential equations. Employing the
factorization procedure, the solutions of the latter equations are
divided into the two classes of bosonic (nonsingular) and fermionic
(singular) cosmological solutions. We next introduce a coupling
parameter denoted by $\rm K$ between the two classes of solutions
and obtain barotropic cosmologies with dissipative features acting
on the scale factors and spatial curvature of the universe. The $\rm
K$-extended FRW equations in comoving time are presented in explicit
form in the low coupling regime. The standard barotropic FRW
cosmologies correspond to the dissipationless limit $\rm K =0$.
}{}{}
%%%%%%%%%%%%%%%%%%%%%%%%%%%

%\centerline{{\tiny written for the Annual Essay Competition of the Gravity Research Foundation for the year 2005}}

%\vspace*{10pt}
%\keywords{The contents of the keywords}

\textlineskip                  %) USE THIS MEASUREMENT WHEN THERE IS
\vspace*{12pt}                 %) NO SECTION HEADING

\vspace*{1pt}\textlineskip  %) USE THIS MEASUREMENT WHEN THERE IS
%\section{General Appearance}    %) A SECTION HEADING
\vspace*{-0.5pt}
\noindent

%%%%%%%%%%%%%%%%%%%%%%%%%%%%%%%%%%%%%%%%%%%%%%
%PACS number(s):  98.80.Hw, 11.30.Pb

\noindent
%%%%%%%%%%%%%%%%%%%%%%%%%%%%%%%%%%%%%%%%%%%%%%%%%%%%%%%%%%%%%%%%%%%%%

\bigskip

%%111111111111111111111111111111111111111111111111111111111111111111111111
%%%%%%%%%%%%%%%%%%%%%%%%%%%%%%%%%%%
%%%%%%%%%%%%%%%%%%%%%%%%%%%%%%
\underline{{\em Comoving FRW barotropy}}

\medskip

\noindent
%The barotropic FRW cosmologies obey the following set of differential equations:
%The scale factor ${\rm a(t)}$ of a FRW metric is a function of the

\noindent Barotropic FRW cosmologies in comoving time ${\rm t}$ obey
the Einstein-Friedmann dynamical equations for the scale factor
${\rm a(t)}$ of the universe supplemented by the (barotropic)
equation of state of the cosmological fluid
%................................................
$$
%\left\{
\begin{array} {lll}
\rm  \frac{\ddot{a}}{a}=-\frac{4\pi G}{3}(\rho +3p)~, \\
%\eqno(1)
%\end{equation}
%\begin{equation} \label{e2}
\rm  H_{0}^{2}(t)\equiv\left(\frac{\dot{a}}{a}\right)^2=\frac{8\pi G\rho}{3}-\frac{\kappa}{a^2}~,\\
%\eqno(2)
%\end{equation}
%\begin{equation} \eqno{e3}
\rm p=(\gamma -1)\rho~,
%\eqno(3)
\end{array}
%\right.}
$$
%.................................................
where $\rho$ and ${\rm p}$ are the energy
density and the pressure, respectively,
of the perfect fluid of which a classical universe is usually assumed to be
made of, $\kappa=0,\pm1$ is the curvature index of the flat, closed, open
universe, respectively, and $\gamma$ is the constant adiabatic index of the cosmological fluid.

%\newpage
%%2222222    2222222222222     222222222222222      222222222222222222          22222222222222222
%%%%%%%%%%%%%%%%%%%%%%%%%%%%%%
%%%%%%%%%%%%%%%%%%%%%%%%%%

\medskip

\underline{{\em Bosonic FRW barotropy}}

\medskip

\noindent
%Combining the equations (1)-(3) and using
Passing to the conformal time variable $\eta$, defined through
${\rm dt=a(\eta)d\eta}$, one can combine the three equations in a single Riccati equation for the Hubble parameter %one
%............................................
%\begin{equation} \label{comb}
%{\rm \frac{a^{''}}{a}+(c-1)\left(\frac{a^{'}}{a}\right)^2+c\kappa=0~.}
%\eqno(5)
%\end{equation}
%.............................................
%where ${\rm c=\frac{3}{2}\gamma -1}$.
%The case $\kappa=0$ is directly integrable,\cite{Far} and will be skipped henceforth.
%The logarithmic derivative
${\rm H_0}(\eta)$ (we shall use either $\rm \frac{d}{d \eta}$ or $'$ for the derivative with respect to $\eta$ in the following)
%..................................................
\begin{equation} \label{ricc}
{\rm H^{'}_{0}+cH^2_{0}+\kappa c=0~,}
%\eqno(6)
\end{equation}
%....................................................
where ${\rm c=\frac{3}{2}\gamma -1}$.

\medskip

Employing now ${\rm H_0(\eta)=\frac{1}{c}\frac{w^{'}}{w}}$ one gets the
very simple (harmonic oscillator) second order differential equation
%.............................................
\begin{equation} \label{w}
{\rm w^{''}-c\cdot c_{\kappa,b}w=0~,}
%\eqno(7b)
\end{equation}
%............................................
where ${\rm  c_{\kappa,b}=-\kappa c}$.
Moreover,  the particular Riccati solutions
%${\rm u_{p}=\frac{1}{c}\frac{w^{'}}{w}}$ mentioned above is only the particular solution, i.e.,
${\rm H_{0}^{+}=-\tan c\eta}$ and ${\rm H_{0}^{-}={\rm coth} \,c\eta}$
for $\kappa =\pm 1$, respectively, are related to the common factorizations of the equation (\ref{w})
%................................................
\begin{equation} \label{w3}
{\rm \left(\frac{d}{d\eta}+cH_{0}\right)
\left(\frac{d}{d\eta}-cH_{0}\right)w=
w^{''}-c(H_{0}^{'}+cH_{0}^{2})w=0}~.
%\eqno(7c)
\end{equation}
%.................................................

%............................................................
%\begin{equation} \label{schr1}
%{\rm w^{''}+\kappa c^2w=0~.}
%\eqno(7a)
%\end{equation}
%......................................................
%For $\kappa =1$
Borrowing a terminology from supersymmetric quantum mechanics, we call the solutions $\rm w$ as bosonic zero modes. They are the following.
For $\kappa =1$
$$
{\rm w_{1,b} \sim \cos(c\eta +d)}\qquad \rightarrow \qquad {\rm a_{1,b}(\eta) \sim {\rm w_{1}^{1/c}}  % [\cos(c\eta +d)]^{1/c}
~,}
$$
where ${\rm d}$ is an arbitrary phase,
whereas for $\kappa =-1$ one gets
$$
{\rm w_{-1,b} \sim {\rm sinh}(c\eta)}\qquad \rightarrow \qquad {\rm a_{-1,b}(\eta) \sim {\rm w_{-1}^{1/c}}    %[{\rm sinh}(c\eta)]^{1/c}
}~.
$$
%where ${\rm W_{\pm 1}}$ and ${\rm A_{\pm 1}}$ are amplitude parameters.
%Eqs. (8) and (9)
%According to Faraoni, the alternative method requires $\gamma$ be a
%constant.

\bigskip

\underline{{\em Fermionic FRW barotropy}}

\medskip

\noindent
A class of barotropic FRW cosmologies with inverse scale factors with respect to the bosonic ones can be obtained  by considering the supersymmetric partner (or fermionic)
equation of Eq.~(\ref{w3}) which is obtained by applying the factorization brackets in reverse order
%...........................................
\begin{equation} \label{f}
{\rm
\left(\frac{d}{d\eta}-cH_{0}\right)
\left(\frac{d}{d\eta}+cH_{0}\right)w=
w^{''}-c(-H_{0}^{'}+cH_{0}^2)w= 0}~.
\end{equation}
Thus, one can write
%...........................
\begin{equation}\label{f1}
{\rm w^{''}
-c\cdot c_{\kappa, f}w=0}~,
%\eqno(7d)
\end{equation}
%.........................................
where
%..........................................
$$
{\rm c_{\kappa, f}(\eta)=-H_{0}^{'}+cH_{0}^2=
\left\{ \begin{array}{ll}
c(1+2{\rm tan}^2 c\eta) & \mbox{if $\kappa =1$}\\
c(-1+2{\rm coth}^2 c\eta) & \mbox{if $\kappa =-1$}
\end{array} \right.}
%\eqno(7d)
$$
%...........................................
%\end{equation}
denotes the supersymmetric partner adiabatic index of fermionic type associated through the mathematical scheme to the constant bosonic index.
Notice that the fermionic adiabatic index is time dependent.
The fermionic $\rm w$ solutions are
$$
{\rm w_{1,f} =\frac{c}{\cos (c\eta +d)}} \qquad \rightarrow \qquad {\rm a_{1,f}(\eta) \sim [\cos(c\eta +d)]^{-1/c}}~,
$$
and
$$
{\rm w_{-1,f} =\frac{c}{sinh (c\eta)}}\qquad \rightarrow \qquad  {\rm a_{-1,f}(\eta) \sim [{\rm sinh}(c\eta)]^{-1/c}}~,
$$
for $\kappa =1$ and $\kappa =-1$, respectively.

\medskip

We can see that the bosonic and fermionic barotropic cosmologies are reciprocal to each other, in the sense that
$$
{\rm a_{\pm,b}a_{\pm,f}=const}~.
$$
Thus, bosonic expansion corresponds to fermionic contraction and viceversa.

\bigskip

%**********************************************************************************
\underline{{\em Uncoupled fermionic and bosonic FRW barotropies}}

\medskip

%**************************************************************************************
\noindent
A matrix formulation of the previous results is possible as follows.
%%33333333333333333333333333333333333333333333333333333333333333333
%%%%%%%%%%%%%%%%%%%%%%%%
Introducing the following two Pauli matrices
$$\alpha =-{\rm i}\sigma _{{\rm y}}=-{\rm i}\left( \begin{array}{cc}
0 & -{\rm i} \\
{\rm i} & 0\end{array} \right ) \qquad {\rm and} \quad
\beta =\sigma _{{\rm x}}=\left( \begin{array}{cc}
0 & 1\\
1 & 0 \end{array} \right )
$$
we can write a cosmological matrix equation
%.................................................
\begin{equation} \label{HD}
%{\rm {\cal H}_{0}^{{\rm FRW}}W=
{\rm \sigma _y D_{\eta}W+\sigma _x (icH_0)W}=0~,
\end{equation}
%....................................................
where ${\rm W}=\left( {\rm \begin{array}{cc}
w_1\\
w_2\end{array}} \right ) $ is a two component `zero-mass' spinor. %$\left(\begin{array} &\Psi _1 & \Psi _2 \end{array} &
This is equivalent to the following decoupled equations
%..............................................
\begin{eqnarray}
{\rm D}_{\eta}w_1+{\rm cH_0}w_1=0 \\                                     % -P_{\eta}w_1+icH_0w_1=0\\
-{\rm D}_{\eta}w_2+{\rm cH_0}w_2=0~.                                                                            %+P_{\eta}w _2+icH_0w_2=0~.
\end{eqnarray}
%............................................
Solving these equations one gets $w_1\propto 1/\cos ({\rm c}\eta)$ and $w_2\propto \cos({\rm c}\eta)$ for $\kappa =1$ cosmologies
and $w_1\propto 1/{\rm sinh} ({\rm c}\eta)$ and $w_2\propto {\rm sinh}({\rm c}\eta)$ for $\kappa =-1$ cosmologies.
Thus, we obtain
$$
W=\left( {\rm \begin{array}{cc}
w_1\\
w_2\end{array}} \right )=\left( {\rm \begin{array}{cc}
{\rm w_f}\\
{\rm w_b}\end{array}} \right )~.
$$
This shows that the matrix equation contains the two reciprocal barotropic cosmologies on the same footing as the two components of the spinor $\rm W$.

%************************************************************************************************************

\newpage

\medskip

\underline{{\em Coupled fermionic and bosonic cosmological
barotropies}}

\medskip

\noindent
There is a simple way to couple the two spinorial components by means of
a constant parameter ${\rm K}$. Indeed, we write
%%%%%%%%%%%%%%%%%%%%
\begin{equation} \label{HDM}
%{\rm H_{K}^{FRW}W=
{\rm [\sigma _y D_{\eta}+\sigma _x (icH_0 +K)]}W={\rm K}W~,
\end{equation}
%%%%%%%%%%%%%%%%%%%%%%%%%%%%%%%%%
where $\rm K$ is equivalent to the mass parameter of a Dirac spinor. Eq.~(\ref{HDM})
is equivalent to the following system of coupled equations
%%%%%%%%%%%%%%%%%%%
\begin{eqnarray}
{\rm D_{\eta}w _1+(i{\rm cH_0+K})w _1={\rm K}w _2}\\
-{\rm D_{\eta}w _2+({\rm icH_0+K})w _2={\rm K}w _1}~.
\end{eqnarray}
%%%%%%%%%%%%%%%%%%%%
These two coupled first-order equations are equivalent to second order differential equations for each of the two spinor components.

\noindent
The fermionic spinor component can be found directly as solutions of
%.....................................
\begin{equation} \label{comp1}
\left\{ \begin{array}{ll}
%{\rm D^{2}_{\eta}
{\rm w_1^{+''}-c[c_{1,f}(\eta)+2iK\tan c\eta] w_1^{+}=0} & \qquad {\rm for} \, \kappa =1\\
%\end{equation}
%.....................................
%and
%%%%%%%%%%%%%%%%%%%%
%\begin{equation} \label{comp1b}
%{\rm D^{2}_{\eta}
{\rm w_1^{-''}-c[c_{-1,f}(\eta)-2iK{\rm coth} \,c\eta] w_1^{-}=0} & \qquad  {\rm for} \, \kappa =-1~,
\end{array} \right.
\end{equation}
%%%%%%%%%%%%%%
whereas the bosonic components are solutions of
%%%%%%%%%%%%%%%%%%%%
\begin{equation} \label{comp2}
\left\{ \begin{array}{ll}
%{\rm D^{2}_{\eta}
{\rm w_2^{+''}+c[c-2iK\tan c\eta] w_2^{+}=0}  & \qquad {\rm for} \quad  \kappa =1\\
%\end{equation}
%%%%%%%%%%%%%%
%and
%...........................................
%\begin{equation} \label{comp2b}
%{\rm D^{2}_{\eta}
{\rm w_2^{-''}+c[-c+2iK{\rm coth} \,c\eta] w_2^{-}=0} & \qquad {\rm for} \quad \kappa =-1~.
\end{array} \right.
\end{equation}
%...........................................
The solutions of the bosonic equations are expressed in terms of the Gauss hypergeometric functions ${\rm _2F_1}$ of complex parameters that can be written in explicit form :
%although we do not enclose them here.
%in the variables ${\rm y=e^{ic\eta}}$ and ${\rm y=e^{c\eta}}$, respectively
%...................................................

\begin{eqnarray} % EQUATION 14
{\rm z_{2}^{-k_2}w_{2}^{+}(\eta)={\cal
A}\,z_{1}^{k_1} %z_{2}^{(q-\frac{1}{2})}
\,_{2} F_{1}\left[k_1+k_2+1,k_1+k_2,1+2k_1\,;-\frac{z_{1}}{2}\right]}\nonumber\\
- {\rm {\cal B}\,e^{-i(1+2k_1)\pi}\left(\frac{4}{z_1}\right)^{k_1}
%%%%z_{2}^{(q-\frac{1}{2})}
\, _{2}F_{1}\left[-k_1+k_2,
-k_1+k_2+1,1-2k_1\,;-\frac{z_{1}}{2}\right]} %e^{i(1-2p)\pi}
\end{eqnarray}
%.............................................
and
%..............................................
\begin{eqnarray} %EQUATION 15
{\rm z_{4}^{-k_4} w_{2}^{-}(\eta) = {\cal C}\,z_{3}^{k_3} \,
_{2}F_{1}\left[k_3+k_4,k_3+k_4+1,1+2k_3;\frac{z_{3}}{2}\right]} \nonumber\\
 + {\rm {\cal D}\,\left(\frac{4}{z_3}\right)^{k_3}\, %z_{3}^{-r} \,
_{2}F_{1}\left[-k_3+k_4+1,-k_3+k_4,1-2k_3;\frac{z_{3}}{2}\right]}~,
%\left[\frac{1}{2} z_{3}\right]^{-2r}
\end{eqnarray}
%..........................................
where the variables ${\rm z}_{i}$ ($i=1,...,4$) are given in the
following form:
$${\rm z_{1}=i\tan({\rm c}\eta )-1, \quad z_{2}=i\tan({\rm c}\eta)+1,\quad
z_{3}={\rm coth}(c \eta )+1, \quad z_{4}={\rm coth}(c \eta)-1},
$$
respectively. The $\rm k$ parameters are the following:
%................................................................
$$
{\rm k_1 %p-\frac{1}{2}
=\frac{1}{2}\left(1-\frac{2K}{c}\right)^{\frac{1}{2}}, \quad k_2
%q-\frac{1}{2}
=\frac{1}{2}\left(1+\frac{2K}{c}\right)^{\frac{1}{2}}},
$$
and
$$
{\rm k_3=\frac{1}{2}\left(1+i\frac{2K}{c}\right)^{\frac{1}{2}},
\quad k_4=\frac{1}{2}\left(1-i\frac{2K}{c}\right)^{\frac{1}{2}}},
$$
whereas ${\cal A}$, ${\cal B}$, ${\cal C}$, ${\cal D}$ are
superposition constants. Plots of these modes are given in Ref.~2b.

\medskip

\noindent
Based on these ${\rm K}$ zero-modes, we can introduce bosonic scale factors and Hubble parameters depending on the parameter ${\rm K}$
\begin{equation}\label{hw+}
{\rm a_{K,+} = (w_2^{+})^{1/c} ~, \qquad \qquad H^{+}_{K}(\eta)=\frac{1}{ c}\left(\log w_2^{+}\right)^{'}}
\end{equation}
and
\begin{equation}\label{hw-}
{\rm a_{K,-} = (w_2^{-})^{1/c} ~, \qquad \qquad H^{-}_{K}(\eta)=\frac{1}{c}\left(\log w_2^{-}\right)^{'}}~,
\end{equation}
and similarly for the fermionic components by changing ${\rm w_2^{\pm}}$ to ${\rm w_1^{\pm}}$ in eqs.~(\ref{hw+})
and (\ref{hw-}), respectively.

\bigskip

\underline{{\em Comoving ${\rm K}$-coupled FRW barotropy: Small ${\rm K}$ regime}}

\medskip

\noindent
%In the small ${\rm K}$ limit, ${\rm K\rightarrow 0}$, the ordinary FRW barotropic cosmologies are obtained.
%The Hubble parameters corresponding to the $\rm K$- dependent bosonic modes are plotted in the Figures~(1) -(4).

\noindent
Introducing the notations ${\rm \lambda _K=-K(\frac{\partial a_K}{\partial K})_{K=0}}$ and ${\rm F_{\kappa}(t)=\left(1+\frac{2\lambda _K}{\kappa a_K}\right)}$,
one can show that in the small K/c limit the comoving time equations can be written as follows:

%................................................
$$
\begin{array}{lll}
%\begin{equation} \label{efinal1}
{\rm \frac{\ddot{a}_K}{a_K}\left(F_1(t)-\frac{\lambda _K}{a_K}\right)-\frac{\ddot{\lambda} _K}{a_K}=-\frac{4\pi G}{3}(\rho +3p)}~,\\
%\eqno(1)
%\end{equation}
%.................................................
%\begin{equation} \label{efinal2}
{\rm
%H_{K}^{2}(t)\equiv
\left(\frac{\dot{a}_K}{a_K}\right)^2F_{1}(t)
%\left(1+\frac{2\lambda _K}{a_K}\right)
=\frac{8\pi G\rho}{3}-\frac{(\kappa  + 2\frac{\lambda _K}{a_K})}{a_{K}^{2}}
%\left(1 +\frac{2\lambda _K}{\kappa a_K}
%-2\dot{a}_K\dot{\lambda}_K
%\right)
%F_{\kappa}(t)
~,}\\
%\eqno(2)
%\end{equation}
%..............................................
%\begin{equation} \label{efinal3}
{\rm p=(\gamma -1)\rho}~.
%\eqno(3)
%\end{equation}
\end{array}
$$
%.................................................
($\rm a_K$ could be either $\rm a_{K,+}$ or $\rm a_{K,-}$ depending on the $\kappa$ case we take into account).

%where
%..................................
%\begin{equation}\label{last}
%{\rm \lambda _K=-K\frac{\partial a_K}{\partial K}|_{K=0}}
%\end{equation}
%..................................
%and
%..........................................
%\begin{equation}\label{last1}
%{\rm \kappa(t;K)=\dot{\kappa}-2\dot{a}_K\dot{\lambda} + (\dot{\lambda})^2}
%\end{equation}
%...........................................

%\newpage
\bigskip
%%44444444444444444444444444444444444444444444444444444444444444444444444444444444444
%%%%%%%%%%%%%%%%%%%%
%%%%%%%%%%%%%%%%%%%%%%
%{\bf 4 - Conclusions}
\underline{{\em Interpretation}}

\medskip

\noindent
We come now to the interpretational issue. We consider only the small $\rm K$ regime as realistic. Then, the
effects of $\rm K$ show up only on the geometrical quantities without any change in the barotropic equation of state.
The parameter $\rm K$ introduces an imaginary part in the cosmological Hubble parameter $\rm H$. Since the latter
is the logarithmic derivative
of the scale factor of the universe one comes to the conclusion that the supersymmetric techniques presented
here are a supersymmetric way to take into
account dissipation and instabilities of barotropic FRW cosmologies. However, since $\rm K$ is also a coupling
parameter between fermion and
boson components, the dissipation and instabilities belong to the cosmological epochs that occurred before
the supersymmetry breaking.

%We recall that in the Newtonian setting, Casti {\em et al} \cite{casti} studied negative
%energy modes and gravitational instability of interpenetrating fluids. An extension of their
%work to relativistic cosmology...

\bigskip

%\nonumsection{Acknowledgements}

%\noindent
%This work was supported by a Project from CONACyT (No.~).

%\newpage

\end{document}